\documentclass[10pt,conference]{IEEEtran}
\usepackage{epsfig,graphicx,subfigure,psfrag,amsmath,cases,bm}
\usepackage{latexsym,amssymb,algorithm,mathtools}
\usepackage{algorithmic}
\usepackage{color}
\usepackage{url}
\usepackage{scrtime}
\usepackage{stfloats}
\usepackage{tablefootnote}
\usepackage{cite}
\usepackage{amsfonts,mathabx}
\usepackage{textcomp}
\usepackage{xcolor,capt-of}
\usepackage{multirow}

\usepackage[left=0.625in,top=0.72in,bottom=0.93in,right=0.625in]{geometry}
\newtheorem{T-Prob}{Transformed Problem}

\newcounter{TempEqCnt}

\DeclareMathOperator{\mino}{minimize}

\DeclareMathOperator{\subto}{subject\hspace*{2mm}to}

\newcommand{\QED}{\hfill \ensuremath{\blacksquare}}

\allowdisplaybreaks

\ifCLASSINFOpdf
\else
\fi
\usepackage{amsmath}

\begin{document}
\title{Joint BS Selection, User Association, and Beamforming Design for Network Integrated Sensing and Communication\vspace*{-2mm}}
\author{Yiming Xu, Dongfang Xu, Lei Xie, and Shenghui Song\\
    {Dept. of ECE, The Hong Kong University of Science and Technology, Hong Kong}\\
    {Email: yxuds@connect.ust.hk, $\{$eedxu, eelxie, eeshsong$\}$@ust.hk}\vspace*{-4mm}}
\maketitle
\begin{abstract}
Different from conventional radar, the cellular network in the integrated sensing and communication (ISAC) system enables collaborative sensing by multiple sensing nodes, e.g., base stations (BSs). However, existing works normally assume designated BSs as the sensing nodes, and thus can't fully exploit the macro-diversity gain. In the paper, we propose a joint BS selection, user association, and beamforming design to tackle this problem. The total transmit power is minimized while guaranteeing the communication and sensing performance measured by the signal-to-interference-plus-noise ratio (SINR) for the communication users and the Cram\'{e}r-Rao lower bound (CRLB) for location estimation, respectively. An alternating optimization (AO)-based algorithm is developed to solve the non-convex problem. Simulation results validate the effectiveness of the proposed algorithm and unveil the benefits brought by collaborative sensing and BS selection.
\end{abstract}

\IEEEpeerreviewmaketitle
\section{Introduction}
Integrated sensing and communication (ISAC) is recognized as a very promising technique for the sixth-generation (6G) wireless networks \cite{liu2020joint} to provide sensing and communication services with shared infrastructure, spectrum resources, and signal processing modules. For this purpose, many appealing results have been obtained regarding the beamforming design, post-signal processing, and resource allocation in ISAC systems \cite{hua2023optimal, liu2020joint, xu2022robust}. One standing challenge for ISAC is the severe self-interference (SI) \cite{xu2023coordinatedisac, liu2022joint, 9933894}. Specifically, the typical round-trip time for radar echo is much shorter than the communication frames and thus the echoes are rebounded to the transmitter before the information transmission ends, causing SI. In addition, the severe attenuation of radar echo caused by the round-trip path loss makes the effect of SI more significant.
\par
To address the severe SI issue in ISAC, the authors of \cite{liu2022joint} and \cite{9933894} investigated the beamforming design of the mono-static system by exploiting the spatial angle-of-arrival (AoA) difference between the SI and the signal of interest (SoI). Another way to avoid SI is to explore the network structure of the ISAC system and assign different base stations (BSs) as the transmitting and receiving sensing nodes, respectively \cite{10077114}.
Along this line of research, the authors of \cite{Behdad2022,huang2022coordinated, demirhan2023cell} investigated the network ISAC system where distributed transmitters send the dual-functional radar and communication (DFRC) signals to serve the communication users (CUs) and locate the sensing target (ST) at the same time. In particular, the authors of \cite{Behdad2022} considered the power allocation problem under the cloud radio access network (C-RAN) architecture, where each access point (AP) serves as either a transmitter or a receiver for sensing purposes. In \cite{huang2022coordinated}, the authors assumed dedicated sensing receivers, where a joint power control method was proposed to minimize the total transmit power. \cite{demirhan2023cell} considered a network ISAC system with distributed transmitting and receiving APs, and jointly designed the sensing and communication beamformers to maximize the signal-to-noise ratio (SNR) for sensing while satisfying the signal-to-interference-plus-noise ratio (SINR) requirement for communication.
\par
The above works obtained engaging results in tackling SI by network ISAC, but all assumed designated sensing transceivers, without fully exploiting the macro-diversity provided by the distributed sensing nodes. To this end, the authors of \cite{xu2023coordinatedisac} proposed a heuristic BS selection algorithm where the receiving BS was selected based on its distance from the CUs and the ST. Then, the transmit and receive beamformers were jointly designed to minimize the beam pattern mismatch error while satisfying the quality-of-service (QoS) requirements by both sensing and communication.
\par
In this paper, we jointly optimize the BS selection, user association, and beamforming strategy in network ISAC systems, considering the influence of sensing on communication, i.e., the BSs selected as sensing receivers can no longer serve CUs for the downlink transmission. The total transmit power is minimized while guaranteeing the communication and sensing performance, where SINR of the CUs and the Cram\'{e}r-Rao lower bound (CRLB) for location estimation are adopted as the communication and sensing metric, respectively. An alternating optimization (AO)-based algorithm is proposed to solve the non-convex problem. Numerical simulations validate  the effectiveness of the proposed method and illustrate the benefits of exploiting the macro-diversity of the network ISAC system.
\par
\textit{Notation:} 
Vectors and matrices are denoted by boldface lowercase and boldface capital letters, respectively. $\mathbb{R}^{M\times N}$ and $\mathbb{C}^{M\times N}$ represent the space of the $M\times N$ real-valued and complex-valued matrices, respectively. $|\cdot|$ and $||\cdot||_2$ denote the absolute value of a complex scalar and the $l_2$-norm of a vector, respectively. $(\cdot)^T$ and $(\cdot)^H$ stand for the transpose and the conjugate transpose of their arguments, respectively. $\mathbf{I}_{N}$ refers to the $N$ by $N$ identity matrix. $\mathrm{tr}(\mathbf{A})$ and $\mathrm{rank}(\mathbf{A})$ denote the trace and the rank of matrix $\mathbf{A}$, respectively. $\mathbf{A}\succeq\mathbf{0}$ indicates that $\mathbf{A}$ is a positive semidefinite matrix. $\Re\{\cdot\}$ and $\Im\{\cdot\}$ represent the real and imaginary parts of a complex number, respectively. Vectorization of matrix $\mathbf{A}$ is denoted by $\mathrm{vec}(\mathbf{A})$. $\mathbf{A}\otimes\mathbf{B}$ denotes the Kronecker product of two matrices $\mathbf{A}$ and $\mathbf{B}$. $\mathbb{E}[\cdot]$ refers to statistical expectation. $\overset{\Delta }{=}$ and $\sim$ stand for ``defined as'' and ``distributed as'', respectively.

\section{System Model and Problem Formulation}
Consider a network ISAC system with $S$ BSs indexed by $s \in \mathcal{S}\overset{\Delta }{=}\left\{1,...,S\right\}$, $K$ communication users (CUs) indexed by $k \in \mathcal{K}\overset{\Delta }{=}\left\{1,...,K\right\}$, and one target (ST). Each BS is equipped with a uniform linear array (ULA) comprising $N$ antennas while all the CUs are assumed to be single-antenna devices. BSs are connected to a central controller to facilitate synchronization and joint design.
To explore the macro-diversity for sensing while minimizing the influence on communication service, we propose to select one BS as the receiving BS and the others as the transmitting BSs.
\begin{figure}[t]
\centering
\includegraphics[width=3.0in]{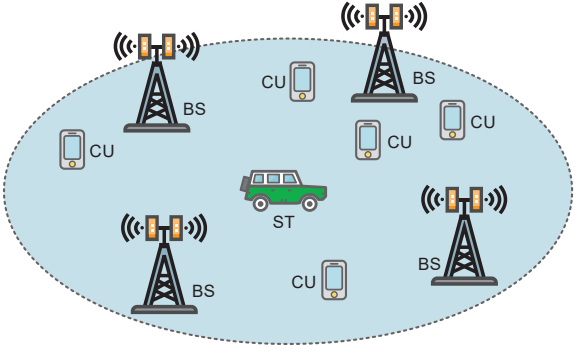}
\vspace*{-2mm}
\caption{Illustration of the considered network ISAC system comprising $S=4$ base stations (BSs), $K=5$ communication users (CUs), and one interesting sensing target (ST).}
\label{System}
\vspace*{-4mm}
\end{figure}
\subsection{Transmit Signal}
The transmitting BSs send DFRC signals for both sensing and communication purposes \cite{hua2023optimal}.
In the time slot $l \in \mathcal{L}:=\left[1,...,L\right]$, the transmit signal of the $s$-th BS is given by
\vspace*{-1mm}
\begin{equation}
\mathbf{x}_s[l] = \underset{k\in\mathcal{K}}{\sum} a_{s,k}\mathbf{w}_{s,k}s_{s,k}[l] + b_s\mathbf{s}_{s,0}[l],\\[-1mm]
\end{equation}
where $a_{s,k}\in\{0,1\}$ is the BS-CU association variable. Specifically, $a_{s,k}=1$ implies that $k$-th CU is served by the $s$-th BS and vice versa. $b_s\in\{0,1\}$ is the BS selection variable. In particular, $b_s=1$ indicates  that the $s$-th BS is selected as a transmitting BS while $b_s=0$ means it serves as the receiving BS.
$\mathbf{w}_{s,k} \in \mathbb{C}^{N}$ and $\mathbf{s}_{s,0}[l] \in \mathbb{C}^{N}$ denote the beamforming vector from the $s$-th BS to the $k$-th CU and the sensing beamformer of the $s$-th BS at the $l$-th time slot, respectively. $s_{s,k}[l]$ denotes the communication signal from the $s$-th BS to the $k$-th CU at the $l$-th time slot, where the signals at different time slots are assumed to be independent and identically distributed (i.i.d.) Gaussian random variables with zero mean and unit variance. The dedicated sensing signal $\mathbf{s}_{s,0}[l]$ is independent of the communication signals. The sample covariance matrix of the transmitted signal from the $s$-th BS is given by
\vspace*{-1mm}
\begin{equation}
\mathbb{E}\big(\mathbf{x}_s[l]\mathbf{x}_s^H[l]\big) = \underset{k\in\mathcal{K}}{\sum} a_{s,k}\mathbf{w}_{s,k}\mathbf{w}^H_{s,k}+ b_s \mathbf{R}_s,\\[-1mm]
\end{equation}
where $\mathbf{R}_s$ denotes the sample covariance matrix of the sensing signal given by
\vspace*{-2mm}
\begin{equation}
\mathbf{R}_s = \frac{1}{L}\underset{l \in \mathcal{L}}{\sum} \mathbf{s}_{s,0}[l] \mathbf{s}^H_{s,0}[l].\\[-2mm]
\end{equation}

\subsection{Communication Model}
The received signal of the $k$-th CU at the $l$-th time slot is given by
\vspace*{-2mm}
\begin{eqnarray}
y_{\mathrm{c},k}[l]&\hspace*{-2mm}=\hspace*{-2mm}& \underset{s\in\mathcal{S}}{\sum}\underset{k'\in\mathcal{K}}{\sum} a_{s,k'} \mathbf{h}_{s,k}^H \mathbf{w}_{s,k'}s_{s,k'}[l]\notag\\
&\hspace*{-2mm}+\hspace*{-2mm}&\underset{s\in\mathcal{S}}{\sum} b_s\mathbf{h}_{s,k}^H\mathbf{s}_{s,0}[l]+z_{\mathrm{c},k}[l],\\[-2mm]\notag
\end{eqnarray}
where $\mathbf{h}_{s,k} \in \mathbb{C}^{N} $ denotes the channel from the $s$-th BS to the $k$-th CU. $z_{\mathrm{c},k}[l] \sim \mathcal{CN}(0,\sigma_{\mathrm{c},k}^2)$ represents the additive white Gaussian noise (AWGN) at the $k$-th CU. The channel state information (CSI) of all communication channels is assumed to be known via proper channel estimation methods \cite{zheng2022survey}. 

The signal-to-interference-plus-noise ratio (SINR) at the $k$-th CU is given by
\vspace*{-5mm}
\begin{equation} \label{SINR}
\hspace*{-2mm}\gamma_{k}=\frac{\underset{s\in \mathcal{S}}{\sum}a_{s,k}\left \vert \mathbf{h}_{s,k}^H\mathbf{w}_{s,k} \right \vert^2 }{\hspace*{-1mm}\underset{s\in\mathcal{S}}{\sum}\big(\underset{k'\in\mathcal{K}\setminus\{k\}}{\sum}\hspace*{-1mm}a_{s,k'}\left \vert \mathbf{h}_{s,k}^H\mathbf{w}_{s,k'} \right \vert^2\hspace*{-1mm}+b_{s} \mathbf{h}_{s,k}^H \mathbf{R}_{s}\mathbf{h}_{s,k}\big)\hspace*{-0.5mm}+ \hspace*{-0.5mm}\sigma_{\mathrm{c},k}^2}.\\[-1mm]
\end{equation}
\subsection{Sensing Model}
We consider the sensing task to locate a point target. If the $s$-th BS is selected as the receiving BS (all other BSs will serve as the transmitting BS), its received echo is given by
\vspace*{-1mm}
\begin{equation}
\mathbf{y}_{\mathrm{r},s}[l] = \underset{s'\in\mathcal{S}\setminus\{s\}}{\sum}\alpha_{s,s'}\mathbf{G}_{s,s'}\mathbf{x}_{s'}[l-\tau_{s,s'}] + \mathbf{z}_{\mathrm{r},s}[l],\\[-1mm]
\end{equation}
where $\alpha_{s,s'}$ is the complex gain containing the target's
radar cross section (RCS) and the path loss of the link from the
$s'$-th BS to the ST and then to the $s$-th BS. $\mathbf{G}_{s,s'}\overset{\Delta}{=}\mathbf{a}(\theta_{\mathrm{r},s})\mathbf{a}^H(\theta_{\mathrm{t},s'})$ is the target response matrix where $\mathbf{a}(\theta) = [1, e^{-j\pi \sin \theta},...,^{-j(N-1)\pi \sin \theta}]^T$ denotes the steering vector. $\theta_{\mathrm{r},s}$ and $\theta_{\mathrm{t},s'}$ represent the angle-of-arrival (AoA) and angle-of-departure (AoD) of the ST with respect to the receiving and transmitting BS, respectively. $\tau_{s,s'}$ denotes the time delay of signals transmitted from the $s'$-th BS to the ST and then reflected to the $s$-th BS. $z_{\mathrm{r},s}[l] \sim \mathcal{CN}(0,\sigma_{\mathrm{r},s}^2)$ represents the AWGN at the $s$-th BS.
As a lower bound for the variance of unbiased estimators, CRLB is adopted as the sensing performance metric and evaluated in the following proposition.
\newtheorem{prop}{Proposition}
\begin{prop}
Assuming the $s$-th BS is selected as the receiving BS, the corresponding CRLB for estimating the ST's location denoted by the two-dimensional coordinate $\mathbf{p}=[p_1, p_2]^T$ is given by
\vspace*{-2mm}
\begin{equation} \label{CRLB}
\mathrm{CRLB}^{(s)}(\mathbf{p}) = \mathrm{tr}\left( \left(\mathbf{F}_{\mathbf{p}\mathbf{p}}- \mathbf{F}_{\mathbf{p}\bm{\alpha}_s} \mathbf{F}_{\bm{\alpha}_s\bm{\alpha}_s}^{-1} \mathbf{F}_{\mathbf{p}\bm{\alpha}_s}^T\right)^{-1} \right),\\[-1mm]
\end{equation}
\end{prop}
where $\mathbf{F}_{\mathbf{p}\mathbf{p}}$, $\mathbf{F}_{\mathbf{p}\bm{\alpha}_s}$ and $\mathbf{F}_{\bm{\alpha}_s\bm{\alpha}_s}$ are sub-matrices of the Fisher information matrix (FIM) defined in \eqref{fisher_matrix1}. 
\par
\textit{Proof:} Please refer to the Appendix. \QED
\subsection{Problem Formulation}
The objective of this paper is to minimize the total transmit power while guaranteeing the QoS requirements for both communication and sensing. The joint BS selection, user association, and beamforming design, i.e., $b_s$, $a_{s,k}$, $\mathbf{w}_{s,k}$, and $\mathbf{R}_s$, is obtained by solving the following optimization problem
\vspace*{-1mm}
\begin{eqnarray}
\label{prob1}
&&\hspace{-9mm}\underset{\substack{\mathbf{R}_s\in\mathbb{H}^{N},\mathbf{R}_s\succeq \mathbf{0},\\\mathbf{w}_{s,k},a_{s,k},b_s}}{\mino} \,\, \,\, \hspace*{0mm}\underset{s\in\mathcal{S}}{\sum}\underset{k\in\mathcal{K}}{\sum}a_{s,k}\left\| \mathbf{w}_{s,k} \right\|_2^2+\underset{s\in\mathcal{S}}{\sum}b_s\mathrm{tr}(\mathbf{R}_s)\notag\\
&&\hspace*{-8mm}\subto\hspace*{1mm}\mbox{C1:}\hspace*{1mm}\underset{k\in\mathcal{K}}{\sum}a_{s,k}\left\Vert\mathbf{w}_{s,k} \right\Vert_2^2 + b_s\mathrm{tr}(\mathbf{R}_s) \leq P,\hspace*{1mm}\forall s,\notag \\ 
&&\hspace*{10mm}\mbox{C2:}\hspace*{1mm}\underset{s\in\mathcal{S}}{\sum}{(1-b_s)\mathrm{CRLB}^{(s)}(\mathbf{p})} \leq \epsilon,\notag \\ 
&&\hspace*{10mm}\mbox{C3:}\hspace*{1mm}\gamma_{k} \geq \Gamma_{k}, \hspace*{1mm} \forall k,\hspace*{2mm}\mbox{C4:}\hspace*{1mm}\underset{s\in\mathcal{S}}{\sum} a_{s,k} = 1,\hspace*{1mm}\forall k,\notag\\
&&\hspace*{10mm}\mbox{C5:}\hspace*{1mm}\underset{s\in\mathcal{S}}{\sum} (1-b_{s}) = 1,\hspace*{1mm}\mbox{C6:}\hspace*{1mm}b_s \geq a_{s,k},\hspace*{1mm}\forall s, \forall k,\notag\\
&&\hspace*{10mm}\mbox{C7:}\hspace*{1mm}a_{s,k}\hspace*{-0.5mm}\in\hspace*{-0.5mm}\{0,1\}, \forall s, \forall k,\hspace*{1mm}\mbox{C8:}\hspace*{1mm}b_s\hspace*{-0.5mm}\in\hspace*{-0.5mm}\{0,1\}, \forall s.\\[-6mm]\notag
\end{eqnarray}
Here, the real-valued scalar $P$ in constraint C1 denotes the maximum transmit power of each BS. A predefined threshold for the CRLB, $\epsilon$, is introduced in constraint C2. Similarly, we impose a predefined minimum SINR requirement for each CU via $\Gamma_k$ in constraint C3. Each CU is assumed to be served by only one BS, as specified in constraints C4 and C7. Constraints C5 and C8 ensure that only one BS can be selected as the receiving BS. Constraint C6 guarantees that only the transmitting BSs can be assigned to serve CUs for the downlink transmission.
\section{Joint Optimization}
Note that problem \eqref{prob1} is a non-convex combinatorial optimization problem which is known to be NP-hard. In particular, the non-convexity originates from the coupling between the optimization variables, the fractional function in constraint C3, and the binary constraints C7 and C8. In fact, there is no systematic algorithm that can optimally solve problem \eqref{prob1} in polynomial time. In this section, we develop a low-complexity suboptimal algorithm by exploiting the AO theory.
\vspace{-1mm}
\subsection{Reformulation of the SINR and CRLB Constraints}
First, we reformulate the SINR constraint C3 and the CRLB constraint C2 for the convenience of solving the the problem.
Define the beamforming matrix $\mathbf{W}_{s,k}\in\mathbb{C}^{N\times N}$ as $\mathbf{W}_{s,k}\overset{\Delta}{=}\mathbf{w}_{s,k}\mathbf{w}_{s,k}^H$, $\forall s$. We can then rewrite constraint C3 equivalently as
\vspace*{-3mm}
\begin{eqnarray}
\overline{\mbox{C3}}\mbox{:}\hspace*{1mm}&&\hspace*{-6mm}\frac{1}{\Gamma_{k}} \underset{s\in \mathcal{S}}{\sum}a_{s,k}\mathrm{tr}(\mathbf{H}_{s,k}\mathbf{W}_{s,k}) -\sum_{s\in \mathcal{S}} b_{s} \mathrm{tr}(\mathbf{H}_{s,k} \mathbf{R}_{s})\notag\\ 
&&\hspace*{-8mm}- \underset{s\in\mathcal{S}}{\sum}\underset{k'\in\mathcal{K}\setminus\{k\}}{\sum}a_{s,k'}\mathrm{tr}( \mathbf{H}_{s,k}\mathbf{W}_{s,k'}) \geq \sigma_{z_k}^2, \hspace{1mm}\forall k,\\[-3mm]\notag
\end{eqnarray}
where $\mathbf{H}_{s,k}\in\mathbb{C}^{N\times N}$ is defined as $\mathbf{H}_{s,k}\overset{\Delta}{=}\mathbf{h}_{s,k}\mathbf{h}_{s,k}^H$, $\forall s,k$.
\par
Next, we reformulate the CRLB constraint C2 into a more tractable form. In particular, we note that $\mathbf{F}_{\mathbf{x}\mathbf{x}}^{(s)}- \mathbf{F}_{\mathbf{x}\bm{\alpha}}^{(s)} \big[\mathbf{F}_{\bm{\alpha}\bm{\alpha}}^{(s)}\big]^{-1} \big[\mathbf{F}_{\mathbf{x}\bm{\alpha}}^{(s)}\big]^T \succeq \mathbf{0}$ and the function $\mathrm{tr}\left(\mathbf{C}^{-1}\right)$ is decreasing on the positive
semidefinite matrix space. By introducing an auxiliary optimization variable $\mathbf{J} \in \mathbb{C}^{2 \times 2}$, \eqref{CRLB} can be equivalently expressed as the following three constraints
\vspace*{-1mm}
\begin{eqnarray}
&&\hspace*{-12mm}\mbox{C2a:}\hspace*{1mm}\mathrm{tr}(\mathbf{J}^{-1}) \leq \epsilon,\hspace*{6mm}\mbox{C2b:}\hspace*{1mm}\mathbf{J} \succeq \mathbf{0}, \\
&&\hspace*{-12mm}\mbox{C2c:}\hspace*{0mm}\sum_s(1-b_s)\hspace*{-0.5mm}\left(\mathbf{F}_{\mathbf{x}\mathbf{x}}^{(s)} -\mathbf{F}_{\mathbf{x}\bm{\alpha}}^{(s)} \big[\mathbf{F}_{\bm{\alpha}\bm{\alpha}}^{(s)}\big]^{-1} \big[\mathbf{F}_{\mathbf{x}\bm{\alpha}}^{(s)}\big]^T\hspace*{-1mm}-\mathbf{J} \right)\hspace*{-0.5mm}\succeq \mathbf{0}.\\[-4mm]\notag
\end{eqnarray}
Moreover, by exploiting the Schur complement \cite{zhang2006schur}, constraint C2c can be rewritten equivalently as
\vspace*{-1mm}
\begin{equation}
\overline{\mbox{C2c}}\mbox{:} \hspace*{1mm} \sum_s(1-b_s)
\begin{bmatrix}
        \mathbf{F}_{\mathbf{x}\mathbf{x}}^{(s)}-\mathbf{J} & \mathbf{F}^{(s)}_{\mathbf{x}\bm{\alpha}}\\
        \big[\mathbf{F}^{(s)}_{\mathbf{x}\bm{\alpha}}\big]^{T} & \mathbf{F}^{(s)}_{\bm{\alpha}\bm{\alpha}}\\
\end{bmatrix}
\succeq \mathbf{0}.
\end{equation}
Now, problem \eqref{prob1} is reformulated into an equivalent form as
\vspace*{-1mm}
\begin{eqnarray}
\label{prob2}
&&\hspace{-6mm}\underset{\substack{\mathbf{W}_{s,k},\mathbf{R}_s, \mathbf{J}\in\mathbb{H}^{N},\\\mathbf{W}_{s,k},\mathbf{R}_s\succeq \mathbf{0},a_{s,k},b_s}}{\mino} \,\, \,\, \hspace*{0mm} f \overset{\Delta}{=} \underset{s\in\mathcal{S}}{\sum}\underset{k\in\mathcal{K}}{\sum}a_{s,k}\mathrm{tr}(\mathbf{W}_{s,k})+\underset{s}{\sum}b_s\mathrm{tr}(\mathbf{R}_s)\notag\\
&&\hspace*{-2mm}\subto\hspace*{6mm}\mbox{C1},\mbox{C2a},\mbox{C2b},\overline{\mbox{C2c}},\overline{\mbox{C3}},\mbox{C4-C8},\notag\\
&&\hspace*{21mm}\mbox{C9:}\hspace*{1mm}\mathrm{rank}(\mathbf{W}_{s,k})\leq 1,\hspace*{1mm}\forall s,\hspace*{1mm}\forall k.\\[-6mm]\notag
\end{eqnarray}
Here, $\mathbf{W}_{s,k}\in\mathbb{H}^{N}$, $\mathbf{W}_{s,k}\succeq \mathbf{0}$, and the unit-rank constraint C9 are imposed to guarantee that we can recover a feasible beamforming vector $\mathbf{w}_{s,k}$ for the original problem \eqref{prob1} after optimization. In the next subsection, we tackle the non-convex binary constraints C7 and C8.
\subsection{Tackling the Binary Constraints C7 and C8}
To handle the binary optimization variables, we first rewrite constraints C7 and C8 equivalently as follows
\vspace*{-1mm}
\begin{eqnarray}
&&\hspace*{-11mm}\mbox{C7a:}\hspace*{1mm} \underset{s\in\mathcal{S}}{\sum}\underset{k\in\mathcal{K}}{\sum}(a_{s,k}-a_{s,k}^2)\leq 0,\hspace*{1mm}\mbox{C7b:}\hspace*{1mm} 0\leq a_{s,k} \leq 1,\forall s, \forall k,\\
&&\hspace*{-11mm}\mbox{C8a:}\hspace*{1mm} \underset{s}{\sum}(b_s-b_s^2)\leq 0,\hspace*{12.3mm}\mbox{C8b:}\hspace*{1mm} 0\leq b_s \leq 1,\forall s.\\[-5mm]\notag
\end{eqnarray}
Here, constraints C7a and C7b are non-convex. To circumvent this issue, we employ the penalty method \cite{nocedal1999numerical} and introduce a penalty term to the objective function in \eqref{prob2} as follows
\vspace*{-1mm}
\begin{equation}
f + \mu\underset{s\in\mathcal{S}}{\sum}\underset{k\in\mathcal{K}}{\sum}\left(-a_{s,k}^2 -b_s^2 + a_{s,k} + b_s\right),\\[-1mm]
\end{equation}
where $\mu\gg 1$ is a penalty factor to avoid any infeasible solution for the binary constraints C7 and C8. Subsequently, we apply the majorization-minimization technique to convexify this non-convex penalty term. In particular, for given feasible points $a^{(t)}_{s,k}$ and $b^{(t)}_{s}$ obtained from the $t$-th iteration of the algorithm\footnote{Note superscript $t$ denotes the iteration index of \textbf{Algorithm 1}, which will be introduced in the next subsection.}, we construct a tractable convex surrogate function $\overline{f}$ for the original objective function $f$ by employing the first-order Taylor expansion on $a_{s,k}^2+b_s^2$. In particular, the convex surrogate function $\overline{f}$ is defined as
\vspace*{-2mm}
\begin{eqnarray}
\overline{f}\overset{\Delta}{=}f&\hspace*{-2mm}+\hspace*{-2mm}&\mu\underset{s\in\mathcal{S}}{\sum}\underset{k\in\mathcal{K}}{\sum}\left( \big(1-2a^{(t)}_{s,k} \big)a_{s,k}\right.\notag\\
&\hspace*{-2mm}+\hspace*{-2mm}& \left.\big(1-2b^{(t)}_{s}\big)b_{s} + \big(a^{(t)}_{s,k} \big)^2  + \big(b_s^{(t)}\big)^2\right).\\[-5mm]\notag
\end{eqnarray}
Note that function $\overline{f}$ is a linear function with respect to (w.r.t.) all optimization variables.
\subsection{Alternating Optimization-Based Algorithm}
In this subsection, we overcome the coupling between the optimization variables and develop a computationally-efficient suboptimal algorithm. To start with, we employ the big-M method \cite{griva2009linear} to tackle the coupled terms $a_{s,k}\mathbf{W}_{s,k}$ in \eqref{prob2}. In particular, we define a new optimization variable $\widetilde{\mathbf{W}}_{s,k}\in\mathbb{C}^{N\times N}$ where $\widetilde{\mathbf{W}}_{s,k}=a_{s,k}\mathbf{W}_{s,k}$. Then, we further reformulate the relation between $\widetilde{\mathbf{W}}_{s,k}$ and $\mathbf{W}_{s,k}$ into a set of convex constraints as follows
\begin{eqnarray}
&&\mbox{C10a:}\hspace*{1mm}\widetilde{\mathbf{W}}_{s,k} \preceq a_{s,k}P\mathbf{I},\hspace*{1mm}\forall s,\hspace*{1mm}\forall k,\\
&&\mbox{C10b:}\hspace*{1mm}\widetilde{\mathbf{W}}_{s,k} \succeq \mathbf{W}_{s,k}-(1-a_{s,k})P\mathbf{I},\hspace*{1mm}\forall s,\hspace*{1mm}\forall k,\\
&&\mbox{C10c:}\hspace*{1mm}\widetilde{\mathbf{W}}_{s,k} \preceq \mathbf{W}_{s,k},\hspace*{1mm}\forall s,\hspace*{1mm}\forall k,\\
&&\mbox{C10d:}\hspace*{1mm}\widetilde{\mathbf{W}}_{s,k} \succeq \mathbf{0},\hspace*{1mm}\forall s,\hspace*{1mm}\forall k.
\end{eqnarray}
Next, by exploiting the AO theory \cite{bezdek2002some}, we divide the optimization variables into two disjoint blocks, i.e., $\left\{ \mathbf{b} \right\}$ and $\left\{ \mathbf{W}_{s,k}, \widetilde{\mathbf{W}}_{s,k}, \mathbf{R}_s, \mathbf{J}, \mathbf{A} \right\}$, where each block is associated with a subproblem. Then, by fixing one block, we update the other by solving the corresponding subproblem. In particular, in the $(t+1)$-th iteration, the two resulting subproblems are solved in an alternating manner as follows 
\subsubsection{Update $\{\mathbf{b}\}$}
For the given block $\left\{ \mathbf{W}_{s,k}, \widetilde{\mathbf{W}}_{s,k}, \mathbf{R}_s, \mathbf{J}, \mathbf{A} \right\}$, we update $\{\mathbf{b}\}$ by solving the following subproblem\footnote{To simplify the notation, we omit the constant term from the objective function $\overline{f}$ in both subproblems. On the other hand, when checking the convergence condition, i.e., step 6 of the proposed algorithm, we also take into account these constant terms.} 
\vspace*{-2mm}
\begin{eqnarray}
\label{prob3}
&&\hspace{-6mm}\underset{b_s}{\mino} \,\, \,\, \hspace*{2mm}\underset{s}{\sum}b_s\mathrm{tr}(\mathbf{R}_s) + \mu\underset{s\in\mathcal{S}}{\sum}\left(\big(1-2b^{(t)}_{s}\big)b_{s} + \big(b_s^{(t)}\big)^2  \right) \notag\\
&&\hspace*{-7mm}\subto\hspace*{3mm}\mbox{C1},\overline{\mbox{C2c}},\overline{\mbox{C3}},\mbox{C5},\mbox{C6},\mbox{C8}.
\end{eqnarray}
Note that problem \eqref{prob3} is a convex problem w.r.t. $b_s$ and can be solved optimally by convex solvers such as CVX \cite{grant2014cvx}.
\subsubsection{Update $\left\{ \mathbf{W}_{s,k}, \widetilde{\mathbf{W}}_{s,k}, \mathbf{R}_s, \mathbf{J}, \mathbf{A} \right\}$}
After obtaining $\mathbf{b}$, we solve for $\left\{ \mathbf{W}_{s,k}, \widetilde{\mathbf{W}}_{s,k}, \mathbf{R}_s, \mathbf{J}, \mathbf{A} \right\}$ by focusing on the following subproblem
\vspace*{-2mm}
\begin{eqnarray}
\label{prob4}
&&\hspace{-6mm}\underset{\substack{\mathbf{W}_{s,k},\mathbf{R}_s\in\mathbb{H}^{N_{\mathrm{T}}},\widetilde{\mathbf{W}}_{s,k},\\\mathbf{W}_{s,k},\mathbf{R}_s\succeq \mathbf{0}, \mathbf{A}}}{\mino} \,\, \,\, \hspace*{0mm} \underset{s\in\mathcal{S}}{\sum}\underset{k\in\mathcal{K}}{\sum}\mathrm{tr}(\widetilde{\mathbf{W}}_{s,k})+\underset{s}{\sum}b_s\mathrm{tr}(\mathbf{R}_s)\notag\\ &&\hspace{22mm}+ \mu \underset{s \in \mathcal{S}}{\sum} \underset{k \in \mathcal{K}}{\sum} \left( \big(1-2a^{(t)}_{s,k} \big)a_{s,k} + \big(a^{(t)}_{s,k} \big)^2 \right)  
\notag\\
&&\hspace*{0mm}\subto\hspace*{6mm} \mbox{C1},\mbox{C2a},\mbox{C2b},\overline{\mbox{C2c}},\overline{\mbox{C3}},\mbox{C4},\notag\\
&&\hspace*{23mm} \mbox{C6},\mbox{C7},\mbox{C9},\mbox{C10a-C10d}.
\vspace{-5mm}
\end{eqnarray}
The only obstacle for efficiently solving problem \eqref{prob4} is the non-convex rank-one constraint C9. By using the SDR technique, we can remove constraint C9 and the resulting rank-relaxed version of problem \eqref{prob4} can be solved by CVX. The tightness of the relaxation is revealed in the following theorem.
\par
\textit{Theorem 1:}\hspace*{1mm} An optimal beamforming matrix $\mathbf{W}_k$ with unit-rank can always be obtained by solving
problem \eqref{prob4} without constraint C9.
\par
\textit{Proof:}\hspace*{1mm}\textit{Theorem 1} can be proved by following the same steps as in \cite[Appendix A]{yu2020irs}. The detailed
proof is omitted due to space limitations. \QED
\par
The proposed algorithm is summarized in \textbf{Algorithm 1}. Some additional remarks on the algorithm are as follows:
\par
\textit{i) Convergence and optimality:} In each iteration of \textbf{Algorithm 1}, we solve two convex subproblems, i.e., problem \eqref{prob3} and the rank-relaxed version of problem \eqref{prob4}, in an alternating manner. As a result, the objective function $\overline{f}$ is monotonically non-increasing over iterations and hence the proposed algorithm is guaranteed to converge \cite{bezdek2002some}. Moreover, it has been shown in \cite[Ch. 17]{nocedal1999numerical} that for $\varepsilon \to 0$ and $\mu \uparrow +\infty$, the limit of any convergent sequence generated by \textbf{Algorithm 1} is a stationary point of \eqref{prob2}.
\par
\textit{ii) Complexity:} Note that in each iteration of \textbf{Algorithm 1}, we solve problem \eqref{prob3} for a set of scalar variables, while problem \eqref{prob4} involves a number of positive semidefinite matrices. As a result, the computational complexity of \textbf{Algorithm 1} is dominated by step 4, i.e., the optimization of a semidefinite programming (SDP) problem. According to \cite[Theorem 3.12]{polik2010interior}, the per iteration computational complexity of \textbf{Algorithm 1} is given by $\mathcal{O}\Big(\mathrm{log}\frac{1}{\kappa}\big((2SK+S)^2N^2+(2SK+S)N^3\big)\Big)$, where $\mathcal{O}\left ( \cdot  \right )$ is the big-O notation and $\kappa$ is the accuracy factor when employing the interior point method to solve an SDP problem.
\begin{algorithm}[t]
\caption{AO-Based Algorithm}
\begin{algorithmic}[1]
\small
\STATE Set iteration index $t=1$, error tolerance factor $0<\varepsilon\ll1$, and penalty factor $\mu\gg 1$. Initialize the optimization variables $\mathbf{b}^{(1)}$, $\mathbf{W}^{(1)}_{s,k}$, $\widetilde{\mathbf{W}}^{(1)}_{s,k}$, $\mathbf{R}^{(1)}_s$, $\mathbf{J}^{(1)}$, and $\mathbf{A}^{(1)}$
\REPEAT
\STATE Solve problem \eqref{prob3} for given $\mathbf{W}^{(t)}_{s,k}$, $\widetilde{\mathbf{W}}^{(t)}_{s,k}$, $\mathbf{R}^{(t)}_s$, $\mathbf{J}^{(t)}$, and $\mathbf{A}^{(t)}$, and update the solution $\mathbf{b}^{(t+1)}$
\STATE Solve the rank-relaxed version of \eqref{prob4} for given $\mathbf{b}^{(t+1)}$ and update $\mathbf{W}^{(t+1)}_{s,k}$, $\widetilde{\mathbf{W}}^{(t+1)}_{s,k}$, $\mathbf{R}^{(t+1)}_s$, $\mathbf{J}^{(t+1)}$, and $\mathbf{A}^{(t+1)}$
\STATE Set $t=t+1$
\UNTIL $\frac{\left|\overline{f}^{(t)}-\overline{f}^{(t-1)}\right|}{\overline{f}^{(t)}}\leq \varepsilon$
\end{algorithmic}
\end{algorithm}
\section{Simulation Results}
\begin{figure}[t]
\centering
\includegraphics[width=3.2in]{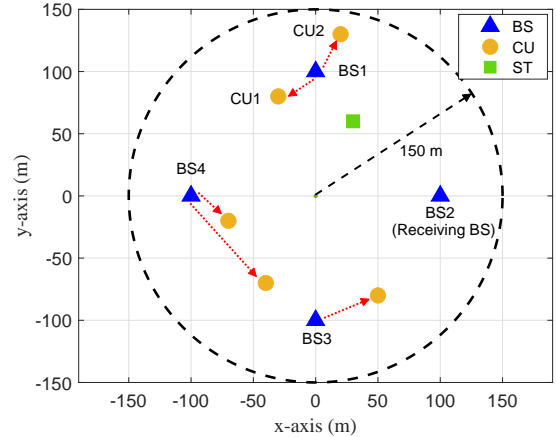}
\vspace*{-2mm}
\caption{Illustration of the considered ISAC system with $S=4$ and $K=5$, and the radius of the service area is $150$ m. By adopting the proposed scheme, BS2 is selected as the receiving BS. The red dashed arrows denote the resulting BS-CU association policy.}
\label{setup}
\vspace*{-4mm}
\end{figure}

\begin{table}[t]\vspace*{0mm}\caption{System Simulation Parameters.\vspace*{-2mm}}\label{parameter}\footnotesize
\newcommand{\tabincell}[2]{\begin{tabular}{@{}#1@{}}#2\end{tabular}}
\centering
\renewcommand{\arraystretch}{1.2}
\begin{tabular}{|l|l|l|}
\hline
    \hspace*{-1mm}$\sigma_{\mathrm{c},k}^2$ & Noise power at CU $k$ & $-85$ dBm \\
\hline
    \hspace*{-1mm}$\sigma_{\mathrm{r},s}^2$ & Noise power at the receiving BS & $-65$ dBm \\
\hline
    \hspace*{-1mm}$P$ & Maximum transmit power at each BS & $45$ dBm \\
\hline
    \hspace*{-1mm}$N$ &  Number of antennas & $8$ \\
\hline
    \hspace*{-1mm}$L$ &  Number of radar samples & $1024$ \\
\hline
    \hspace*{-1mm}$\mu$ & Penalty parameter & $3\times10^{4}$ \\
\hline
    \hspace*{-1mm}$\varepsilon$ & Error tolerance factor & $10^{-3}$ \\
\hline
\end{tabular}
\vspace*{-4mm}
\end{table}
In this section, we evaluate the performance of the proposed scheme via simulation. The system setting is shown in Fig. \ref{setup}. In particular, the service area is assumed to be a circle with a radius of $150$ m. The system comprises $S=4$ BSs, $K=5$ CUs, and one ST. The $4$ BSs are located at $[0\mathrm{m},100\mathrm{m}]$, $[100\mathrm{m},0\mathrm{m}]$, $[0\mathrm{m},-100\mathrm{m}]$, and $[-100\mathrm{m},0\mathrm{m}]$, respectively, while the CUs and the ST are randomly distributed in the concerned area. The path loss exponent of the BS-CU links is set to $3$, while that of the BS-ST link is set to $2$, and Rayleigh fading is assumed for the BS-CU channel. Unless otherwise specified, we adopt the parameters summarized in Table \ref{parameter}.
\par
For comparison purposes, we consider three baseline schemes. Scheme 1 selects the BS closest to the ST as the receiving BS and then jointly optimizes the BS-CU association and beamforming. Scheme 2 performs the receiving BS selection and CU assignment in a random manner and then optimizes the beamforming. Scheme 3 considers the bi-static situation with one transmitting BS and one receiving BS (pre-selected as BS1 and BS2) which is widely adopted to avoid the SI issue. The BS selection, BS-CU association, and beamforming are jointly optimized in Scheme 3. 
\par
To illustrate the BS selection and BS-CU association strategy of the proposed scheme, we first consider a specific setup where $K=5$ CUs are located at $[-30\mathrm{m},80\mathrm{m}]$, $[20\mathrm{m},130\mathrm{m}]$, $[50\mathrm{m},-80\mathrm{m}]$, $[-40\mathrm{m},-70\mathrm{m}]$ and $[-70\mathrm{m},-20\mathrm{m}]$, respectively, and the ST is located at $[30\mathrm{m},60\mathrm{m}]$, cf. Fig. \ref{setup}. 
The parameters are set as $\Gamma_k=8$dB and $\epsilon=1$. Table \ref{BS_selection_policy} illustrates the advantage of the proposed scheme over Scheme 1. As shown in Fig. \ref{setup}, although BS1 is the closest BS to the ST, it is not selected by the proposed scheme as the receiving BS, but instead the transmitting BS to serve two CUs, because of the locations of the two CUs. This underlines the importance of joint BS selection and user association.
\par
\begin{table}[t]\vspace*{0mm}\caption{Comparison of different BS selection policy.\vspace*{-2mm}}\label{BS_selection_policy}\footnotesize
\newcommand{\tabincell}[2]{\begin{tabular}{@{}#1@{}}#2\end{tabular}}
\centering
\renewcommand{\arraystretch}{1.2}
\begin{tabular}{|c|c|c|}
\hline
    \hspace*{-1mm} Method & Receiving BS & Total transmit power (dBm) \\
\hline
    \hspace*{-1mm} Proposed scheme & BS2& $29.7$ dBm \\
\hline
    \hspace*{-1mm} Baseline scheme 1 & BS1& $38.7$ dBm \\
\hline
\end{tabular}
\vspace*{-3mm}
\end{table}
\begin{figure}[t]
\centering
\includegraphics[width=3.2in]{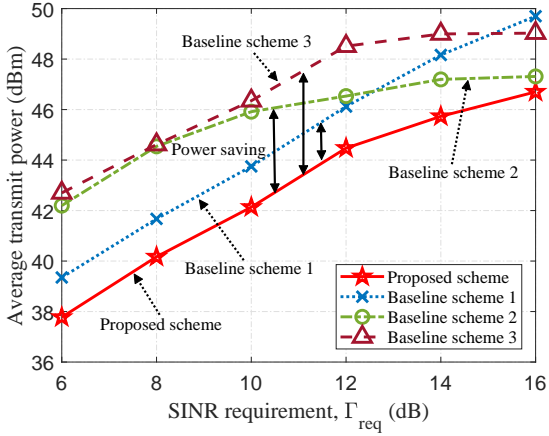}
\vspace*{-2mm}
\caption{Average total transmit power versus minimum SINR requirement.}
\label{figure:power_sinr}
\vspace*{-4mm}
\end{figure}
\begin{figure}[t]
\centering
\includegraphics[width=3.2in]{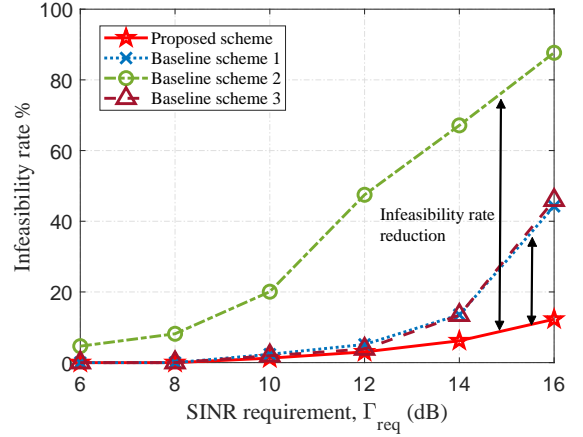}
\vspace*{-2mm}
\caption{Infeasibility rate versus minimum SINR requirement.}
\label{figure:infeasibility_rate_sinr}
\vspace*{-4mm}
\end{figure}
Fig. \ref{figure:power_sinr} depicts the average total transmit power of the system versus the minimum SINR requirement $\Gamma_{\mathrm{req}}$, where $\epsilon$ is set as $1$. We can observe that the proposed scheme outperforms all baseline schemes. The power saving gap between the proposed scheme and Scheme 1 increases with $\Gamma_{\mathrm{req}}$. This is because Scheme 1 adopts the sensing-prioritized strategy which harms the overall performance when the QoS requirement of communication increases. The gap between the proposed scheme and Scheme 3 comes from the macro-diversity gain of network ISAC. Furthermore, we notice that as $\Gamma_{\mathrm{req}}$ increases, the gap between the proposed scheme and Scheme 2 shrinks. This is because infeasibility occurs if a
solution cannot satisfy the QoS requirements with a given power constraint, and the curves in Fig. \ref{figure:power_sinr} are only averaged over the feasible solutions. The infeasibility rate versus $\Gamma_{\mathrm{req}}$ is shown in Fig. \ref {figure:infeasibility_rate_sinr} which shows the advantage of the proposed scheme, and the big gap between the proposed scheme and Scheme 2 comes from selection diversity.
\section{Conclusion}
In this paper, we investigated the benefits of macro-diversity for sensing applications in network ISAC systems.
Different from existing works that assumed designated transmitting and receiving BSs, we proposed a joint BS selection, BS-CU association, and beamforming design scheme to exploit the macro-diversity while taking the influence of sensing on communication into account.
An AO-based algorithm is proposed to solve the non-convex optimization problem in which the total transmit power of the system is minimized while satisfying the QoS requirements of communication and sensing. Simulation results validated the effectiveness of the proposed algorithm and revealed the benefits of collaborative sensing in exploiting the macro-diversity for sensing purposes.
\begin{figure*}[t]
\setcounter{TempEqCnt}{\value{equation}} 		 
	\setcounter{equation}{33}
\begin{eqnarray}
F^{(s)}_{p_ip_j}&\hspace*{-2mm}=\hspace*{-2mm}&\frac{2L}{\sigma_{\mathrm{r},s}^2} \Re\left\{  \left( \underset{s'\in\mathcal{S} \setminus \{s\}}{\sum}\alpha_{s,s'}^* \mathrm{vec}\left(\Tilde{\mathbf{G}}^{(i)}_{s,s'}\mathbf{X}_{s,s'}\right)^H \right) \left( \underset{s'\in\mathcal{S} \setminus \{s\}}{\sum}\alpha_{s,s'} \mathrm{vec}\left(\Tilde{\mathbf{G}}^{(j)}_{s,s'}\mathbf{X}_{s,s'}\right) \right) \right\}\label{F_pp}\notag\\
&\hspace*{-2mm}=\hspace*{-2mm}& \frac{2L}{\sigma_{\mathrm{r},s}^2} \Re\left\{ \mathrm{tr} \left(\underset{s'\in\mathcal{S}\setminus\{s\}}{\sum}\vert\alpha_{s,s'} \vert^2 \Tilde{\mathbf{G}}^{(j)}_{s,s'}\left(\underset{k\in\mathcal{K}}{\sum}a_{s',k}\mathbf{w}_{s',k}\mathbf{w}^H_{s',k} + {\mathbf{R}_{s'}} \right) [\Tilde{\mathbf{G}}^{(i)}_{s,s'}]^H \right) \right\},\hspace*{1mm}i,j \in \{1,2\}.\\
\mathbf{F}^{(s)}_{\mathbf{p}\bm{\alpha}}&\hspace*{-2mm}=\hspace*{-2mm}&\frac{2L}{\sigma_{\mathrm{r},s}^2} \Re\left\{  
    \begin{bmatrix}
        \underset{s'\in\mathcal{S} \setminus \{s\}}{\sum}\alpha_{s,s'}^* \mathrm{vec}\left(\Tilde{\mathbf{G}}^{(1)}_{s,s'}\mathbf{X}_{s,s'}\right)^H \\
        \underset{s'\in\mathcal{S} \setminus \{s\}}{\sum}\alpha_{s,s'}^* \mathrm{vec}\left(\Tilde{\mathbf{G}}^{(2)}_{s,s'}\mathbf{X}_{s,s'}\right)^H \\
    \end{bmatrix} [1,j] \otimes \left[\mathrm{vec}\left(\mathbf{G}_{s,s'}\mathbf{X}_{s,s_1'}\right),...,\mathrm{vec}\left(\mathbf{G}_{s,s'}\mathbf{X}_{s,s_{S-1}'}\right)\right]
\right\}\label{F_pa}\notag\\
&\hspace*{-2mm}=\hspace*{-2mm}&\frac{2L}{\sigma_{\mathrm{r},s}^2} \Re\left\{ [1,j] \otimes \mathbf{D}_{2\times (S-1)}\right\}.\\
\mathbf{D}_{1,s'_i}&\hspace*{-2mm}=\hspace*{-2mm}&\alpha^*_{s,s'_i}\mathrm{tr} \left( \mathbf{G}_{s,s'_i} \left(\underset{k\in\mathcal{K}}{\sum}a_{s'_i,k}\mathbf{w}_{s_i',k}\mathbf{w}^H_{s_i',k} + {\mathbf{R}_{s'_i}}\right)\left[\Tilde{\mathbf{G}}^{(1)}_{s,s'_i}\right]^H \right),\hspace*{1mm}s_i'\in\mathcal{S}\setminus\{s\},\label{D_1s}\\
\mathbf{D}_{2,s'_i}&\hspace*{-2mm}=\hspace*{-2mm}&\alpha^*_{s,s'_i}\mathrm{tr} \left( \mathbf{G}_{s,s'_i} \left(\underset{k\in\mathcal{K}}{\sum}a_{s'_i,k}\mathbf{w}_{s_i',k}\mathbf{w}^H_{s_i',k} + {\mathbf{R}_{s'_i}}\right)\left[\Tilde{\mathbf{G}}^{(2)}_{s,s'_i}\right]^H \right),\hspace*{1mm}s_i'\in\mathcal{S}\setminus\{s\}.\label{D_2s}\\
\setcounter{equation}{39}
\mathbf{E}_{i,i}^{(s)}&\hspace*{-2mm}=\hspace*{-2mm}&\mathrm{tr}\left(\mathbf{G}_{s,s'_i}\left(\underset{k\in\mathcal{K}}{\sum}a_{s'_i,k}\mathbf{w}_{s_i',k}\mathbf{w}^H_{s_i',k} + {\mathbf{R}_{s'_i}}\right)\mathbf{G}_{s,s'_i}^H \right), \hspace*{1mm}s_i'\in\mathcal{S}\setminus\{s\},\hspace*{1mm}i\in\left\{1,\cdots,S-1\right\}.\label{E_ii}
\end{eqnarray}
\setcounter{equation}{25}
\hrule
\end{figure*}
\section*{Appendix \\ Proof of Proposition 1}
Assume that the $s$-th BS is selected as the receiving BS. By stacking the received echoes $\mathbf{y}_{\mathrm{r},s}[l]$, $l=1,\cdots,L$, over $L$ samples, we have
\vspace*{-1mm}
\begin{equation} \label{stacked}
\mathbf{Y}_{\mathrm{r},s} = \underset{s'\in\mathcal{S}\setminus\{s\}}{\sum}\alpha_{s,s'}\mathbf{G}_{s,s'}\mathbf{X}_{s,s'} + \mathbf{Z}_{\mathrm{r},s},
\vspace*{-2mm}
\end{equation}
where $\mathbf{Y}_{\mathrm{r},s}$, $\mathbf{X}_{s,s'}$, and $\mathbf{Z}_{s}\in\mathbb{C}^{N\times L}$ are given by $\mathbf{Y}_{\mathrm{r},s} = \Big[\mathbf{y}_{\mathrm{r},s}[1],...,\mathbf{y}_{\mathrm{r},s}[L]\Big]$, $\mathbf{X}_{s,s'}=\Big[\mathbf{x}_{s'}[1-\tau_{s,s'}],...,\mathbf{x}_{s'}[L-\tau_{s,s'}]\Big]$, and $\mathbf{Z}_{s}=\Big[z_{\mathrm{r},s}[1],...,z_{\mathrm{r},s}[L]\Big]$, respectively. By vectorizing \eqref{stacked}, we have
\begin{equation} \label{echo}
\tilde{\mathbf{y}}_{\mathrm{r},s} = \mathrm{vec}(\mathbf{Y}_{\mathrm{r},s})= \Tilde{\mathbf{u}}_s + \Tilde{\mathbf{z}}_{\mathrm{r},s},
\end{equation}
where vectors $\Tilde{\mathbf{u}}_s\in\mathbb{C}^{NL\times 1}$ and $\Tilde{\mathbf{z}}_{\mathrm{r},s}\in\mathbb{C}^{NL\times 1}$ are given by $\Tilde{\mathbf{u}}_s = \underset{s'\in\mathcal{S}\setminus\{s\}}{\sum}\alpha_{s,s'}\mathrm{vec}(\mathbf{G}_{s,s'}\mathbf{X}_{s,s'})$ and $\Tilde{\mathbf{z}}_{\mathrm{r},s} \sim \mathcal{CN}(\mathbf{0},\mathbf{M}_{\mathrm{r},s})$ with $\mathbf{M}_{\mathrm{r},s} = \sigma^2_{\mathrm{r},s}\mathbf{I}_{NL}$, respectively.
\par
The unknown parameter set to be estimated is $\bm{\xi}_s = [\mathbf{p}^T, \bm{\alpha}_s^T]^T$, where parameter $\bm{\alpha}_s\in\mathbb{R}^{2(S-1)\times 1}$ is given by $\bm{\alpha}_s=
\Big[\Re\{\alpha_{s,s'_1}\},...,\Re\{\alpha_{s,s'_{S-1}}\}, \Im\{\alpha_{s,s'_1}\},...,\Im\{\alpha_{s,s'_{S-1}}\}\Big]^T$, $s' \in \mathcal{S}\setminus\{s\}$.
\par
According to \cite{kay1993fundamentals}, the element of Fisher information matrix (FIM) $\mathbf{F}^{(s)}$ for estimating $\bm{\xi}_s$ is given by
\begin{equation} \label{fisher_element}
    \mathbf{F}^{(s)}_{i,j}\hspace*{-0.5mm}=\mathrm{tr}\hspace*{-0.5mm}\left(\hspace*{-0.5mm} \mathbf{M}_{r,s}^{-1}\frac{\partial \mathbf{M}_{r,s}}{\partial \bm{\xi}_i}\mathbf{M}_{r,s}^{-1}\frac{\partial \mathbf{M}_{r,s}}{\partial \bm{\xi}_j}\hspace*{-0.5mm}\right) + 2\Re\left\{\frac{\partial \Tilde{\mathbf{u}}_s^H}{\partial \bm{\xi}_i}\mathbf{M}_{r,s}^{-1} \frac{\partial \Tilde{\mathbf{u}}_s}{\partial \bm{\xi}_j}\right\}.
\end{equation}
The FIM $\mathbf{F}^{(s)}$ can be partitioned as
\begin{equation} \label{fisher_matrix1}
    \mathbf{F}^{(s)} = \begin{bmatrix}
        \mathbf{F}^{(s)}_{\mathbf{p}\mathbf{p}} & \mathbf{F}^{(s)}_{\mathbf{p}\bm{\alpha}}\\
        \big[\mathbf{F}^{(s)}_{\mathbf{p}\bm{\alpha}}\big]^T & \mathbf{F}^{(s)}_{\bm{\alpha}\bm{\alpha}}\\
    \end{bmatrix}.
    \vspace*{-1mm}
\end{equation}
In the following, we determine the sub-matrices of $\mathbf{F}^{(s)}$ according to \eqref{fisher_element}.
\par
For that purpose, we first calculate the derivatives of $\Tilde{\mathbf{u}}_s$ with respect to the unknown parameters in $\bm{\xi}$, which are given by
\vspace*{-1mm}
\begin{eqnarray}    
\label{u_p1}
\frac{\partial \Tilde{\mathbf{u}}_s}{\partial p_i} &\hspace{-2mm}=\hspace{-2mm}& \hspace*{-1mm}\underset{s'\in\mathcal{S} \setminus \{s\}}{\sum}\hspace*{-0.5mm}\alpha_{s,s'} \mathrm{vec}\left(\dot{\mathbf{G}}^{(i)}_{s,s'}\mathbf{X}_{s,s'} + \mathbf{G}_{s,s'} \dot{\mathbf{X}}_{s,s'}^{(i)}\right),\\
\frac{\partial \Tilde{\mathbf{u}}_s}{\partial \alpha_{s,s'}} &\hspace{-2mm}=\hspace{-2mm}& \mathrm{vec}\left(\mathbf{G}_{s,s'}\mathbf{X}_{s,s'}\right),
\end{eqnarray}
where $i\in \{1,2\}$, and matrices $\dot{\mathbf{G}}^{(i)}_{s,s'} \overset{\Delta }{=}\dot{\mathbf{a}}^{(i)}(\theta_{r,s}) \mathbf{a}^H(\theta_{t,s'}) + \mathbf{a}(\theta_{r,s}) [\dot{\mathbf{a}}^{(i)}(\theta_{t,s'})]^H$ and $\dot{\mathbf{X}}_{s,s'}^{(i)} \overset{\Delta }{=} -\dot{\tau}^{(i)}_{s,s'}\mathbf{X}_{s,s'}$. Here, $\dot{\mathbf{G}}^{(i)}_{s,s'}$, $\dot{\mathbf{X}}_{s,s'}^{(i)}$, $\dot{\mathbf{a}}^{(i)}$ and $\dot{\tau}_{s,s'}^{(i)}$ denote the derivatives of the corresponding term with respect to $p_i$.
By defining $\Tilde{\mathbf{G}}^{(i)}_{s,s'} = \dot{\mathbf{G}}^{(i)}_{s,s'} - \dot{\tau}^{(i)}_{s,s'} \mathbf{G}_{s,s'}$, we can rewrite \eqref{u_p1} as follows
\vspace*{-1mm}
\begin{equation} \label{u_p2}
\frac{\partial \Tilde{\mathbf{u}}_s}{\partial p_i} = \underset{s'\in\mathcal{S} \setminus \{s\}}{\sum}\alpha_{s,s'} \mathrm{vec}\left(\Tilde{\mathbf{G}}^{(i)}_{s,s'}\mathbf{X}_{s,s'}\right), \hspace*{1mm} i \in \{1,2\}.
\vspace{-2mm}
\end{equation} 
\par
We further partition the sub-matrix $ \mathbf{F}^{(s)}_{\mathbf{p}\mathbf{p}}$ in \eqref{fisher_matrix1} as
\vspace*{0mm}
\begin{equation} \label{fisher_matrix2}
    \mathbf{F}^{(s)}_{\mathbf{p}\mathbf{p}} = \begin{bmatrix}
        F^{(s)}_{p_1 p_1} & F^{(s)}_{p_1 p_2}\\
        F^{(s)}_{p_1 p_2} & F^{(s)}_{p_2 p_2}\\
    \end{bmatrix}.
    \vspace{-2mm}
\end{equation}
The elements $F^{(s)}_{p_i p_j}$ can be calculated based on \eqref{fisher_element}, and is given by \eqref{F_pp} at the top of this page.
\par
The sub-matrix $\mathbf{F}^{(s)}_{\mathbf{p}\bm{\alpha}}$ is derived in \eqref{F_pa} according to \eqref{fisher_element}. The matrices $\mathbf{D}_{1,s'_i}$ and $\mathbf{D}_{2,s'_i}$ in \eqref{F_pa} are given by \eqref{D_1s} and \eqref{D_2s}, respectively. Similar to the derivation in \eqref{F_pa}, $\mathbf{F}^{(s)}_{\bm{\alpha}\bm{\alpha}}$ can be obtained as
\vspace*{-2mm}
\setcounter{equation}{37}
\begin{equation}
\mathbf{F}^{(s)}_{\bm{\alpha}\bm{\alpha}} = \frac{2L}{\sigma_{\mathrm{r},s}^2} \mathbf{I}_{2}\otimes \mathbf{E}_{(S-1)\times(S-1)}^{(s)},
\vspace{-2mm}
\end{equation}
where $\mathbf{E}^{(s)}$ is a diagonal matrix with the diagonal element $\mathbf{E}_{i,i}^{(s)}$ given by \eqref{E_ii}, shown at the top of this page.
\par
Given \eqref{fisher_matrix1}, the CRLB for location estimation can be derived as in \eqref{CRLB} according to \cite{bekkerman2006target}. \QED

\bibliographystyle{IEEEtran}
\bibliography{IEEEabrv,reference}
\end{document}